\documentclass[11pt,twoside]{article}
\usepackage[english]{babel}
\usepackage{amssymb}
\usepackage[mathscr]{eucal}
\usepackage{eufrak}
\usepackage{amsmath,amsthm}
\usepackage{mathrsfs}
\usepackage[colorlinks=true,
   urlcolor=blue,           filecolor=green,      
   citecolor=green,      
   linkcolor=red,           bookmarks=true,
  unicode,
   plainpages=false,   ]{hyperref}
\usepackage{color}

\def\bp{\begin{proof}}
\def\ep{\end{proof}}
\def\n{\nabla}
\def\ssfrac#1#2{\mbox{\large$\frac{#1}{#2}$}}
\def\sfrac#1#2{\mbox{\Large$\frac{#1}{#2}$}}
\def\intl#1{\int\limits_{#1}}
\def\intll#1#2{\int\limits_{#1}^{#2}}
\def\dm{|\hskip-0.05cm|}

\def\OO{\Omega}
\def\displ{\displaystyle}
\def\VSE{\vspace{6pt}\\&\displ }
\def\VS{\vspace{6pt}\\\displ }
\def\rf#1{{\rm(\ref{#1})}}
\def\R{\Bbb R}
\def\N{\Bbb N}

\def\vep{\varepsilon}
\def\be{\begin{equation}}
\def\ba{\begin{array}}
\def\ea{\end{array}}
\def\ee{\end{equation}}
\def\ov{\overline}
\def\po{{\partial\Omega}}

\title{\Large\bf Existence of a weak solution to the Navier-Stokes equations\\enjoying the energy equality} 
\author{Paolo Maremonti and Alfonsina Tartaglione\thanks{Dipartimento di Matematica e Fisica,  
Universit\`{a}\,degli 
Studi\,della\,Campania
``L. Vanvitelli'',\,via\,Vivaldi 43,\newline\null\hskip0.55cm 81100 Caserta,
 Italy.\newline\null\hskip0.55cm
paolo.maremonti@unicampania.it\newline\null\hskip0.55cmalfonsina.tartaglione@unicampania.it\newline\null\hskip0.55cm The  research activity  is performed under the
auspices of   GNFM-INdAM. }}
\date{\today}
\begin{document}
\markboth{\footnotesize\rm    P. Maremonti and A. Tartaglione} {\footnotesize\rm  Existence of a weak solution to the Navier-Stokes equations enjoying the energy equality
} \maketitle
 {\small\bf Abstract} - {\small To date, a  Leray's weak solution to the Navier-Stokes equations enjoys  an energy inequality. We prove the existence of a  weak solution that enjoys the energy equality.  The  existence of the weak  solution is achieved as  limit of solutions to the mollified Navier-Stokes system (Leray's approach). Since no uniqueness result is known for weak solutions we are not able to claim that any weak solution in the sense of Leray-Hopf enjoys the energy equality.}\vskip0.2cm{\small\bf Key words}{\small: Navier-Stokes equations, weak solutions, energy equality.}\par{\small Math. Subject Classif.:  35Q30, 35B45, 76D05.}\par
 \newtheorem{ass}
{\bf Assumption}[section]
\newtheorem{defi}
{\bf Definition}[section]
\newtheorem{tho}
{\bf Theorem}[section]
\newtheorem{rem}
{\bf Remark}[section]
\newtheorem{lemma}
{\bf Lemma}[section]
\newtheorem{coro}
{\bf Corollary}[section]
\newtheorem{prop}
{\bf Proposition}[section]
\renewcommand{\theequation}{\thesection.\arabic{equation}}
\setcounter{section}{0}
\numberwithin{equation}{section}
\section{Introduction}
We consider the  Navier-Stokes initial boundary value problem:
\be\label{NS}\ba{cc}v_t+v\cdot\n v+\n\pi=\Delta v\,,&\n\cdot v=0\,,\mbox{ in }(0,T)\times \OO\,,\VS v=0\mbox{ on }(0,T)\times\po\,,& v=v_0\mbox{ on }\{0\}\times\OO\,,\ea\ee where $\OO\subseteq\R^3$ can be a   bounded or an exterior  domain, whose boundary $\po$ for simplicity 
is assumed smooth, a half-space or the whole space. We set $w_t:=\frac\partial{\partial t}w$ and $w\cdot\n w:= (w\cdot\n) w$. \par We investigate   the existence of a weak solution $v$, in the sense of Leray-Hopf, to problem \rf{NS} enjoying the energy equality in the following form:
\be\label{EE}\dm v(t)\dm_2^2+2\intll 0t\dm \n v(\tau)\dm_2^2d\tau=\dm v_0\dm_2^2\,,\;t >0\,.\ee
In 2D  the result is true for all $t>0$. Instead,  the validity of the \rf{EE} is an open problem in $n$D, $n\geq3$. Actually, limiting ourselves to the three-dimensional case, concerning a weak solution, we are able to prove just  an energy inequality. The following one is in the strong form\footnote{\,The energy inequality \rf{NS-III} is said in strong form in contraposition to the one satisfied {\it a priori} only for $s=0$, called in weak form. This is the case $n$D of a Hopf's weak solution to the IBVP in unbounded domains, where no energy   inequality   is  known in the strong form.}   and is due to Leray (1934) \cite{L}:
\be\label{NS-III}\dm v(t)\dm_2^2+2\intll st\dm \n v(\tau)\dm_2^2d\tau\leq \dm v(s)\dm_2^2\,,\mbox{ for all }t>s\mbox{ and a.e. in } s>0 \mbox{ and for }s=0\,.\ee 
{\it A priori} there is no reason of a  strict inequality in \rf{NS-III}. To date, it appears as a consequence of the ``weakness''\footnote{\,In  Leary's  paper \cite{L}, the possible instants of singularity  are heuristically interpreted as possible phenomena of turbulence in a fluid motion, that are expected in  3D.\par We know, in part also thanks to inequality \rf{NS-III},   that the set $S$ of the possible instants of singularity of a Leray weak solution has at least $\mathcal H^\frac12(S)=0$, being $\mathcal H^a$ the $a$-dimensional Hausdorff measure, \cite{CKN,Sh}.} of a weak solution $v$.   \par In order to state our chief result we introduce some notations and the definition  of a weak solution to problem \rf{NS}.\par By the symbol $\mathscr C_0(\OO)$ we mean the set of functions $\{\varphi:\varphi\in C_0^\infty(\OO)\mbox{ and }\n\cdot\varphi=0\}$.
We set $J^2(\OO):=$completion of $\mathscr C_0(\OO)$ with respect the $L^2$-norm.   The symbol  $J^{1,2}(\OO)$ denotes the completion of $\mathscr C_0(\OO)$ in  $W^{1,2}(\OO)$. \begin{defi}\label{DWS}{\sl For all $v_0\in J^2(\OO)$, a field $v:(0,\infty)\times\OO\to\R^3$ is said a weak solution corresponding to  datum $v_0$ if \begin{itemize}\item[{\rm1)}]for all $T>0$, $v\in L^\infty(0,T;J^2(\OO))\cap L^2(0,T;J^{1,2}(\OO))$,\item[{\rm 2)}]for all $T>0$ and  $t,s\in(0,T)$,  the field $v$
satisfies the integral   equation:
\newline \centerline{$\displ\intll
st\Big[(v,\varphi_\tau)-(\nabla
v,\nabla
\varphi)+(v\cdot\nabla\varphi,v)\Big]d\tau+(v(s),\varphi
(s))=(v(t),\varphi(t))$,} for all $\varphi(t,x)\in C_0^1([0,T)\times\OO)$ with $\n\cdot\varphi(t,x)=0$\,,\item[{\rm 3)}]$\displ\lim_{t\to0}\,(v(t),\varphi)=(v_0,\varphi)\,,\mbox{ for all }\varphi\in\mathscr C_0(\OO)\,.$\end{itemize}}\end{defi} 
We are able to prove
\begin{tho}\label{CT}{\sl  For all $v_0\in J^2(\OO)$ there exists a weak solution $v(t,x)$ to problem \rf{NS} enjoying the energy inequality \rf{NS-III} for all $t>s$ and $s\in [0,\infty)-I$, with    $I\subset (0,\theta)$, $\mu_1(I)=0$ and  $\theta\leq c\dm v_0\dm_2^4$. Moreover, $v$ enjoys the energy equality:  \be\label{E-I}\dm v(t)\dm_2^2+2\intll 0t \dm \n v(\tau)\dm_2^2d\tau=\dm v_0\dm_2^2\,,
\mbox{ for all }t\in (0,\infty)-I\,.\ee  In particular, we get
\be\label{E-II}2\intll0\infty\dm\n v(t)\dm_2^2dt=\dm v_0\dm_2^2\,.\ee
Finally, for all $\varphi\in J^2(\OO)$,    the function $(v(t),\varphi)$ belongs to $ C([0,T))$ and $v(t,x)$ is a right continuous function in the $L^2$-norm in all instants $s$ for which \rf{NS-III} holds.}\end{tho}
The proof of Theorem\,\ref{CT}, that in the present note appears very elementary,  is based on a problem of convergence of a sequence of approximating solutions. For this task, we consider the solutions to the mollified Navier-Stokes problem \rf{NSM}: we follow the Leray approach for the construction of a weak solution. For the convergence of the quoted sequence we need two  results. One is related to Lemma\,\ref{WC}, that is a sort of Lebesgue theorem for the dominate convergence, where  the dominated condition for the sequence is substituted  by  a suitable cutoff of values. This is made by means of a suitable auxiliary function.  The second result is given in Lemma\,\ref{SCG}. The result is related to the convergence in  $L^q(0,T;L^2(\OO))$, $q\in[1,2)$, of Leray's approximating sequence $\{\n v^m\}$ to $\n v$ of the weak solution.  Both the results are part of  paper \cite{CGM-III}. In this note, in union  with the above results we are able to furnish a special auxiliary function to achieve the thesis, see \rf{COF} sect.\ref{FPC}. In connection with  the latter result of strong convergence, we stress that a consequence of Theorem\,\ref{CT} is also the convergence in $L^2(0,T;L^2(\OO))$ of the quoted sequence $\{\n v^m\}$. Although the proof is simple, the result is pointed out with details in Corollary\,\ref{SPC}.
\begin{rem}{\rm The inclusion of inequality \rf{NS-III} stated in Theorem\,\ref{CT} is not pleonastic. Actually, estimate \rf{NS-III} holds for all $t>0$, instead equation \rf{E-I} almost everywhere in $t>0$.}\end{rem}
\begin{rem}{\rm Our weak solution enjoys the partial regularity of a weak solution,  in the sense of the Leray's {\it  th\'{e}or\`{e}me de structure}. The partial regularity result can be achieved either  employing the Leray approach (just working on the weak solution \cite{L}), or working on the approximating sequence as proposed in \cite{Mr-Pl} or \cite{M-ST}. \par We would like to point out that recently in \cite{M-ST}, in the case of a bounded domain $\OO$, the author proves a new bound for the instant $\theta$ furnished in Theorem\,\ref{CT}. More precisely, it is proved that $\theta\in [s,s+ S)$ where $s=0$ if  $R(\dm v_0\dm_2)\leq \frac\pi2$, otherwise 
 $s=c(\OO,\nu)\log P(\dm v_0\dm_2) $, and 
 $S:=1/Q(\dm v_0\dm_2)$, where $R(\dm v_0\dm_2)$,  $P(\dm v_0\dm_2)$  and $Q(\dm v_0\dm_2)$ are  polynomials. For fixed domain $\OO$ and viscosity $\nu$,  the new estimate of $\theta$ is better for large value of $\dm v_0\dm_2$, instead $\theta\leq c\dm v_0\dm_2^4$ is better for small value of $\dm v_0\dm_2$. We recall that for the first time the instant $\theta$  was given   by Leray in \cite{L}.}\end{rem}
 \begin{rem}{\rm Of course, the lack of an uniqueness theorem for weak solutions does not allow us to state the result for all weak solutions. In this connection we also stress that,  if the definition of a Leray's weak solution is the one obtained as limit in the metric of the energy of the sequence $\{v^m\}$ to problem \rf{NSM}, then for the uniqueness of the weak limit we can state that a Leary's weak solution enjoys the energy equality \rf{E-II}. Otherwise, if the  Leray's weak solution is just the one corresponding to Definition\,\ref{DWS} with the energy inequality in strong form, then we are not able to state the same result. \par We point out that in the case of Hopf's weak solution related to the initial boundary value problem in bounded domains, the same argument lines are applicable in order to prove energy equality \rf{E-II}.}\end{rem}
 \begin{rem}{\rm
Beyond its important meaning in fluid dynamics, we think that the result of the Theorem is of some interest for a few  analytic questions too. We try to justify the claim.
\par  The energy inequality \rf{NS-III} in the sense of the strict inequality can be interpret as a consequence of  possible turbulence phenomena. Interesting is the conjecture furnished in paper \cite{C-II}, where the authors support the idea that    the presence of turbulences in fluid motions causes an extra  dissipation, the one that balances the gap in the energy inequality \rf{NS-III}.
 Indeed, they consider as possible an analogy between the gap in \rf{NS-III} and  the ``anomalous dissipation''  conjectured by Onsager\footnote{\,The Onsager conjecture is not part of the goals of this note. We refer to the fundamental results obtained in \cite{BDSV,CET,Sh-I}.} (1949) in \cite{Os} for weak solutions to the Euler equations. Now, in the light of \rf{E-II} a possible reformulation of their conjecture can be given in terms of the discontinuity of  $\dm v(t)\dm_2$ for $t\leq \theta$, but not in terms of further dissipation phenomena. Actually,  the energy equality \rf{E-II}  holds almost everywhere in $t>0$. We conclude the connection to the Euler equations  by saying that the approach used in this note cannot be considered by letting viscosity $\nu$ to 0.
\par The compatibility between the initial datum only in $L^2$   and  the validity of the energy equality for  a weak solution was pursued in the recent papers  \cite{CGM-I,CGM-III,Mi,M-EE}. In particular, the result found in \cite{Mi} suggests that there is no  incompatibility. The result of this note  positively works in this sense. 
\par In \cite{BV} there is a result of non uniqueness   for very weak solutions. This result fails to hold if the very weak solution enjoys an energy inequality of the kind \rf{NS-III} for $s=0$, see \cite{BV}.\par In this connection, we think that the energy equality \rf{E-II} states a further constraint in the setting of examples related to the possible ill posed questions for Navier-Stokes equations.\par Our result is proved by assuming the  force  $f=0$. However, this is made for the sake of brevity. The energy equality also holds in the case of $f\in L^2(0,T;L^2(\OO))$ with no change in the proof\footnote{\,We just stress that, for zero initial datum, our assumption $f\in L^2(0,T;L^2(\OO))$ leads to a subset of the solutions considered in \cite{ABC}. Hence, no comparison  is possible between the two solutions.}. On the wake of the previous motivations  concerning the well posedeness and energy relation, we think that the result of energy equality with $f\ne0$ is again of some interest in the light of the recent paper \cite{Pk}, where the possible regularity of a weak solution is achieved by means of arguments involving the behavior of the energy. } \end{rem}
\section{Some preliminary lemmas}
It is well known that $L^2(\OO)\equiv J^2(\OO)\oplus G^2(\OO)\,,$ where  $G^2(\OO):=\{w\in L^2(\OO):w\equiv \n h \mbox{ with }h\in W_{\ell oc}^{1,2}(\OO)\}\,,$ and by the symbol $P$ we denote the projection  from $L^2(\OO)$ onto $J^2(\OO)$. 
\par By the symbol $-P\Delta$ we mean the Stokes operator defined on $J^{1,2}(\OO)\cap W^{2,2}(\OO)$ with range $J^2(\OO)$.
\begin{lemma}\label{GNM}{\sl There exists a constant $c>0$ such that \be\label{GNM-I}\dm w\dm_\infty \leq c\dm P\Delta w\dm_2^\frac12\dm \n w\dm_2^\frac12\,,\mbox{ for all }w\in J^{1,2}(\OO)\cap W^{2,2}(\OO)\,.\ee}\end{lemma}\bp Estimate \rf{GNM-I} is an inequality of the  Gagliardo-Nirenberg kind, whose right-hand side has the Stokes operator as max order of derivatives. In the case of a bounded domain the result is proved in \cite{KF}, but with a constant depending on the domain. For the proof in more general domains see \cite{MRIII,MRIV}. \ep  \begin{lemma}[Friedrichs's lemma]\label{FR}{\sl Let $\OO$ be a bounded domain, and $\{a_p\}$ an orthogonal basis in $L^2(\OO)$. For all $\vep>0$ there exists  $N\in \N$ such that
$$\dm u\dm_2\leq (1+\vep)\big[\mbox{${\overset{N}{\underset{k=1}\sum}}$}(u,a_k)^2\big]^\frac12+\vep\dm\nabla u\dm_q,\mbox{ for all }u\in W^{1,q}(\OO)\,,$$ provided that $q>\frac65$. }\end{lemma}
\bp This lemma is a generalization of the well known Friedrichs's lemma stated for $q=2$. The proof is given in    \cite{LSU} Ch.II Lemma\,2.4.\ep
The following theorem is crucial for our aims. 
\begin{lemma}\label{WC}{\sl Let $\{h_m(t)\}\subset L^1(0,T)$ be a sequence of   non-negative functions such that $\dm h_m\dm_1\leq M<\infty$, for all $m\in\N$. Also, assume that $h_m(t)\to h(t)$ a.e. in $t\in (0,T)$ with $h(t)\in L^1(0,T)$.   Then we get,\be\label{WC-I}\displ\mbox{for all }\alpha\in(0,\widehat\alpha)\,,\quad\lim_m\intll0Th_m(t)p(\alpha,h_m(t))dt= \intll0Th(t)p(\alpha,h(t))dt\,,\ee and \be\label{WC-IO}\lim_{\alpha\to\widehat\alpha^-}\lim_m\intll0Th_m(t)p(\alpha,h_m(t))dt=\intll0Th(t)dt\,,\ee provided that, for all $\alpha\in(0,\widehat\alpha)$, the function $p(\alpha,r)$ is continuous in $r>0$ and 
$$p(\alpha,r):=\left\{\ba{ll}\!1\,,&\mbox{if }r\in[0,g(\alpha)]\,, \VS \!\mbox{is decreasing, with }\displ\lim_{r\to\infty}p(\alpha,r)=0\,, &\mbox{if }r>g(\alpha)\,,\ea\right.$$ where $g(\alpha)$ denotes a strictly increasing and continuous function with $\displ\lim_{\alpha\to\widehat\alpha^-}g(\alpha)=\infty$.
}\end{lemma}
\bp The proof is not invalidated by assuming that for all $m\in\N$  function $h_m$ is defined on $(0,T)$. We have
$$\ba{ll}\displ\intll0Th_m(t)p(\alpha,h_m(t))dt\hskip-0.2cm&\displ=\intll0T(h_m(t)-h(t))p(\alpha,h_m(t))dt+
\intll0Th(t)p(\alpha,h_m(t))dt\\&=:I_1(\alpha,m)+I_2(\alpha,m)\,.\ea$$ Since $h_m(t)\to h(t)$ a.e. in $t\in(0,T)$, then, for all $\alpha\in(0,\widehat\alpha)$, $p(\alpha,h_m(t))\to p(\alpha,h(t))$ a.e. in $t\in (0,T)$. Since $p(\alpha,h_m(t))\leq1$ for all $t\in(0,T)$, recalling that $h(t)\in L^1(0,T)$, then the following limit holds: \be\label{CRGFIC-XII}\lim_mI_2(\alpha,m)=\intll0Th(t)p(\alpha,h(t))dt\,,\mbox{ for all }\alpha\in(0,\widehat\alpha)\,.\ee Now, we consider $I_1$ that, in our assumptions, is  bounded with a constant independent of $\alpha\in(0,\widehat\alpha)$ and $m\in\N$. Our goal is to prove that $\lim_mI_1(\alpha,m)=0$, for all $\alpha\in(0,\widehat\alpha)$. We point out that, for $\vep\in(0,\widehat\alpha-\alpha)$ and for all  $m\in\N$, $$(0,T)=\{t:h_m(t)\leq g(\widehat\alpha-\vep)\}\cup\{t:g(\widehat\alpha-\vep)<h_m(t)\}=:J_m^1(\vep)\cup J_m^2(\vep)\,.$$ Hence, for all $m\in\N$, we write
$$\ba{ll}\displ I_1(\alpha,m)\hskip-0.2cm&\displ=\intll0T\chi_{J^1_m}(t)(h_m(t)-h(t))p(\alpha,h_m(t)) dt+\intll0T\chi_{J_m^2}(t)(h_m(t)-h(t))p(\alpha,h_m(t))dt\VSE=I^1_1(\alpha,m,\vep)+I^2_1(\alpha,m,\vep)\,.\ea$$
Letting $m\to\infty$ we get  $\chi_{J^1_m}(t)|h_m(t)-h(t)|p(\alpha,h_m(t))\leq |h_m(t)-h(t)| \to 0$ a.e. in $t\in (0,T)$. Recalling that $p(\alpha,h_m(t))\leq1$, we have $|\chi_{J^1_m}(t)(h_m(t)-h(t))p(\alpha,h_m(t))|\leq g(\widehat\alpha-\vep)+h(t)$ for all $t\in (0,T)$. Hence, by virtue of Lebesgue's dominated convergence theorem, for all $\alpha\in(0,\widehat\alpha)$ and  for all $\vep\in(0,\widehat\alpha-\alpha)$, we arrive at 
\be\label{CRGFIC-X}\lim_mI_1^1(\alpha,m,\vep)=0\,.\ee
For the second integral we point out that $\chi_{J^2_m}(t)p(\alpha,h_m(t))\leq p(\alpha,g(\widehat\alpha-\vep))$ for all $t\in(0,T)$. Recalling that the sequence $\{h_m\}$ is bounded in $L^1(0,T)$ and that $h\in L^1(0,T)$,  we deduce 
\be\label{CRGFIC-XI}\Big|I_1^2(\alpha,m,\vep)\Big|\leq cp(\alpha,g(\widehat\alpha-\vep))\,,\ee with $c$ independent of $\alpha\in(0,\widehat\alpha)$, $\vep\in(0,\widehat\alpha-\alpha)$ and  $m\in\N$. Hence, via \rf{CRGFIC-X} and \rf{CRGFIC-XI}, for all $\alpha\in(0,\widehat\alpha)$ and $\vep\in(0,\widehat\alpha-\alpha)$, we arrive at $${\overline{\underset m\lim}}\,|I_1(\alpha,m)|\leq cp(\alpha,g(\widehat\alpha-\vep))\,.$$   By the assumptions,  $p(\alpha,r)$ tends to zero for $r\to\infty$, for all $\alpha\in(0,\widehat\alpha)$, and since $\vep\in(0,\widehat\alpha-\alpha)$ is arbitrary,   we arrive at $$\lim_mI_1(\alpha,m)=0\,.$$ The last one and the limit property \rf{CRGFIC-XII} prove \rf{WC-I}. \par Being $h(t)\in L^1(0,T)$ and $p(\alpha,\rho)\leq1$, and, letting $\alpha\to\widehat\alpha$, $ p(\alpha,\rho)\to1$,   via   Lebesgue's    dominated convergence theorem, then from \rf{WC-I} we get $$\lim_{\alpha\to\widehat\alpha} \intll0Th(t)p(\alpha,h(t))dt=\intll0Th(t)dt\,,$$ that proves \rf{WC-IO}. \ep 
\section{The mollified Navier-Stokes system}
We introduce an auxiliary  Navier-Stokes initial boundary value problem:
\be\label{NSM}\ba{cc}v_t^m+ \n\pi^m-\Delta v^m=-\mathbb J_m[v^m]\cdot\n v^m\,,&\n\cdot v^m=0\,,\mbox{ on }(0,T)\times \OO\,,\VS v^m=0\mbox{ on }(0,T)\times\po\,,& v^m=v^m_0\mbox{ on }\{0\}\times\OO\,,\ea\ee
where $\mathbb J_m[\,\cdot\,]\equiv J_{\frac1m}[\,\cdot\,]$ and $J_{\frac1m}[\,\cdot\,]$ is the Friedrichs (space) mollifier, and we tacitly mean that an extension to zero outside $\OO$ is employed for its definition. The sequence $\{v_0^m\}\subset\mathscr C_0(\OO)$ converges to $v_0$ in $L^2$-norm. \par The following Theorem\,\ref{NSM-T} concerns the existence of solutions to problem \rf{NSM}. The result is well known in literature, the first contribute dates  back to  Leray's paper \cite{L}.   
\begin{tho}\label{NSM-T}{For all $v_0^m\in \mathscr C_0(\OO)$ there exists a unique regular solution to problem \rf{NSM} defined for all $T>0$. In particular, for all $T>0$ and $m\in\N$, we get  \be\label{NSM-I}\ba{c}  v^m\in C^1((0,T);J^{1,2}(\OO))\cap L^2(0,T;W^{2,2}(\OO))\,,\;v_t^m\in L^2(0,T;L^2(\OO))\,,\VS \mbox{for all }\eta>0\,,\quad \dm \n v^m(t)\dm_2\in C^1((\eta,T))\,.\ea\ee  Moreover,   the following properties hold:
\be\label{NSM-O}\hskip-0.4cm\displ\mbox{for\,all\,}\varphi\!\in\! J^2(\OO), \{(v^m(t),\varphi)\}\!\subset\! C((0,T))\mbox{\,is\,an\,equicontinuous\,and\,equibounded\,sequence,} \ee 
\be\label{NSM-II}\dm v^m(t)\dm_2^2+2\intll st\dm\n v^m\dm_2^2d\tau= \dm v^m(s)\dm_2^2\leq \dm v_0\dm_2^2\,,
{\,for\,all\,}t>s\geq0\mbox{\,and\,}m\in\N\,,\ee finally,
\be\label{NSM-III}\dm \n v^m(t)\dm_2^2+\sfrac1{2}\intll\theta t\Big[ \dm P\Delta v^m\dm_2^2+\dm v^m_t\dm_2^2\Big]d\tau\leq  (2c\dm  v_0\dm_2^2)^{-1}\,,\mbox{ for all }t\geq\theta\mbox{  and  } m\in\N\,,\ee where  $\theta= c\dm v_0\dm_2^4 $  with a constant $c$  independent of $t,m$ and $v_0$.}\end{tho}
\bp The result of existence of the sequence $\{(v^m,\pi^m)\}$ was achieved in several papers (see {\it e. g.} \cite{CKN, CF, GM, Hy, L}). We give special attention to the result of   existence and  regularity of a solution $(v^m,\pi^m)$ proved by    employing the well known Faedo-Galerkin method,  both in the case of a bounded domain and in the case of an unbounded domain. In the latter case, one achieves the results by making use of the technique of invading domains     as proposed in \cite{Hy} ({\it e.g.}, one finds  this idea developed   in sect.\,2 of \cite{GM} or in the appendix of \cite{CGM-I}). In particular, one proves \rf{NSM-I}-\rf{NSM-II}. We stress that the mollifier $\mathbb J_m$, employed in the definition of system \rf{NSM}, is the key tool in order  to extend, to all $t>0$, the existence of the regular solution\footnote{\,For the convenience of the reader, the    achieved results mentioned  in the previous comments are proved in Appendix. As far as we know, the proof in Appendix is original being based on the method of successive approximations. However, it does  add no new result to the ones already known.}.  \par We prove estimate\footnote{\,Estimate \rf{NSM-III} is generally given  on the weak solution  $v$. It furnishes the regularity of a weak solution $v$ for $t\geq c\dm v_0\dm_2^4$. This is a result related to the Leray's partial regularity,  called structure theorem of a Leray's weak solution.  Here we furnish the proof on $\{v^m\}$. One gets the same regularity  for the  weak solution $v$ with the instant $\theta$ as endpoint of the regularity interval. } \rf{NSM-III}.  We apply the projection operator $P$ to equation \rf{NSM}$_1$, then, we consider the $L^2$-norm of both sides: $$\dm v_t^m-P\Delta v^m\dm_2^2=\dm P(J_m[v^m]\cdot\n v^m)\dm_2^2\,.$$ Since $(v^m_t,P\Delta v^m)=-\frac12\frac d{dt}\dm \n v^m\dm_2^2$, we get the following
\be\label{A}\sfrac d{dt}\dm \n v^m\dm_2^2+\dm P\Delta v^m\dm_2^2+\dm v^m_t\dm_2^2= \dm P(J_m[v^m]\cdot\n v^m)\dm_2^2\,.\ee
We estimate the right-hand side by means of \rf{GNM-I}:
\be\label{ALM-II}\dm P(J_m[v^m]\cdot\n v^m)\dm_2^2\leq \dm J_m[v^m]\cdot\n v^m\dm_2^2\leq \dm v^m\dm_\infty^2\dm \n v^m\dm_2^2\leq c\dm \n v^m\dm_2^3\dm P\Delta v^m\dm_2\,.\ee
Hence, we deduce
\be\label{NSM-IV}\sfrac d{dt}\dm\n v^m\dm_2^2+\sfrac12\dm P\Delta v^m\dm_2^2+\dm v^m_t\dm_2^2\leq c\dm \n v^m\dm_2^6\,.\ee From energy estimate \rf{NSM-II}, for all $m\in\N$, we deduce the existence of a \be\label{ITH}\theta_m\leq c\dm v_0\dm_2^4\;:\;\dm \n v^m(\theta_m)\dm_2^2\leq (2c\dm v_0\dm_2^2)^{-1}.\ee Actually, if the estimate \rf{ITH}  does not hold, we get
$$\sfrac12\dm v_0\dm_2^2=(2c\dm v_0\dm_2^{2})^{-1}c\dm v_0\dm_2^4<\intll0{c\dm v_0\dm_2^4 }\dm\n v^m(\tau)\dm_2^2d\tau\leq\sfrac12\dm v^m_0\dm_2^2\leq\sfrac12\dm v_0\dm_2^2\,,$$ which is an {\it absurdum}. Since the energy relation furnishes  $\dm \n v^m\dm_2^2\leq \dm v^m\dm_2\dm v^m_t \dm_2\leq \dm v_0\dm_2\dm v^m_t\dm_2$, from \rf{NSM-IV} we obtain \be\label{GEX}\sfrac d{dt}\dm \n v^m\dm_2^2+\dm P\Delta v^m\dm_2^2+\dm v^m_t\dm_2^2<\dm\n v^m\dm_2^4 \dm v_0\dm_2^{-2}(2c\dm v_0\dm_2^2\dm \n v^m\dm_2^2-1)_{\big|t=\theta_m}\leq 0\,.\ee This last, the bound for $\theta_m$ and $\dm \n v^m(\theta_m)\dm_2^2\leq (2c\dm v_0\dm_2^2)^{-1}$  easily lead to \rf{NSM-III}.\ep
\begin{lemma}\label{ACLD}{\sl Let $(v^m,\pi_{v^m})$ the solution to problem \rf{NSM} stated in Theorem\,\ref{NSM-T}. Then the following estimate holds:
\be\label{ALCD-I}\dm v^m(t)\dm_{L^2(|x|>R)}^2\leq  \dm v^m_0\dm_{L^2(|x|>\frac R2)}^2+ c(t)\psi(R)=:{\Psi}(R,v_0) \,\mbox{ for any }t>0, \,R>2\ov R\mbox{ and }m\in\N\,,\ee 
  with $c(t)\in L^\infty(0,T)$ and $\Psi(R,v_0)=o(1)$,} {\rm$\ov R>\texttt{diam}\{\R^3-\ov\OO\}$}.\end{lemma}\bp Estimate \rf{ALCD-I} is proved in several papers concerning the question of the energy inequality \rf{NS-III}. For this reason  we do not give the proof, but, for the interested reader, we quote the proof furnished in  sect. 6.4.1 
of the paper \cite{CGM-I}. However,  almost always    the proofs are based on an idea due to Leray in \cite{L}.\ep
\begin{lemma}\label{SCLD}{\sl For all $T>0$, the sequence $\{v^m\}$  furnished by Theorem\,\ref{NSM-T}   weakly converges to $v\in L^2(0,T;J^{1,2}(\OO))$. Moreover, there exists a subsequence, denoted again by $\{v^m\}$, such that,  for all $\varphi\in J^2(\OO)$, the sequence $\{(v^m(t),\varphi)\}$ converges, uniformly on $(0,T)$, to $(v(t),\varphi)\in C((0,T))$  and     converges strongly to $v$ in $L^2(0,T;L^2(\OO))$  }\end{lemma}
\bp The result is well known, it is a part of the existence theorem of a weak solution enjoying the energy inequality in strong form. Thus,  we omit the details and  we limit ourselves to sketch the idea of the proof.   The weak convergence to $v$ is a consequence of the energy relation \rf{NSM-II}. Moreover, for all $\varphi\in J^2(\OO)$, by virtue of \rf{NSM-I}$_1$, any subsequence $\{(v^m(t),\varphi)\}$   converges uniformly on $(0,T)$, and, since by virtue of \rf{NSM-II}, for all $t>0$,  any subsequence $\{v^m\}$  admits $v$ as weak  limit in $J^{2}(\OO)$, then    $(v(t),\varphi)$ is a continuous function, {\it e.g. }see \cite{Ld}. This is enough to apply   the Friedrichs Lemma\,\ref{FR} for the sequence $\{v^m\}\subset L^2(0,T;L^2(\OO))$ in the way suggested in \cite{Ld}, that is, for all $R>0$,
\be\label{FLL}\intll0T\dm v^m-v\dm_{L^2(\OO\cap B_R)}^2d\tau\leq (1+\vep)\mbox{$\underset{p=1}{\overset N\sum}$}\intll0T(v^m-v,a_p)^2d\tau+\vep \intll0T\dm \n v^m-\n v\dm_2^2d\tau\,. \ee
Hence,  the result of convergence  is immediate in the case of $\OO$ bounded. In the case of $\OO$ unbounded, in accordance with  our assumptions, the result can be proved in the way proposed by Leray in \cite{L}. The  idea of the proof  is  to achieve for the sequence $\{v^m\}$ the conditions of the compactness theorem in $L^p(\R^n)$  for all $t>0$.  We set $$H(R,T):=T\dm v_0\dm_{L^2(|x|>R)}^2+\psi(R)\intll0Tc(t)dt +\intll0T\dm v(t)\dm_{L^2(|x|>R)}^2dt\,.$$ By virtue of estimate \rf{ALCD-I}, $H(R,T)=o(1)$ holds.   Then, for all $T>0$, we get
$$\ba{ll}\displ\intll0T\dm v^m(t)-v(t)\dm_2^2dt\hskip-0.2cm&\displ= \intll0T\dm v^m(t)-v(t)\dm_{L^2(\OO\cap B_R)}^2dt+\intll0T\dm v^m(t)-v(t)\dm_{L^2(|x|>R)}^2dt\\&\displ
\leq 2\intll0T\dm v^m(t)-v(t)\dm_{L^2(\OO\cap B_R)}^2dt+2\dm v_0^m-v_0\dm^2_{L(|x|>\frac R2)}+2H(R,t)\,.\ea
$$Employing  \rf{FLL} for the first term on the right-hand side, and recalling that $\{v_0^m\}$ converges to $v_0$ in the  $L^2$-norm, letting  $m\to\infty$, the first two terms on the right-hand side tend  to zero. Moreover, being   $H(R,T)=o(1)$,   letting $R\to\infty$, we get that the right-hand side  approaches zero. \par  
\ep 
The following lemma was proved for the first time in \cite{CGM-I}.
\begin{lemma}\label{SCG}{\sl Let   $\{v^m\}$ be the sequence furnished by Theorem\,\ref{NSM-T} and let $v$ be the limit whose existence is  ensured by Lemma\,\ref{SCLD}. Then, for all $q\in[1,2)$, the sequence $\{v^m\}$ strongly   converges to $v$   in $L^q(0,T;J^{1,2}(\OO))$.}\end{lemma}
\bp  In order to prove the strong convergence, we initially  prove that $\{P\Delta v^m\}$ is bounded in $L^\frac23(0,T;L^2(\OO))$\footnote{\, An analogous integrability property for  weak solutions has been obtained both in \cite{DF} and in \cite{FGT}. But our proof  is original and very short with regards to the ones of the quoted papers and is furnished on the sequence $\{v^m\}$ (we are not interested in the property of the weak solution).}. We consider \rf{NSM-IV} again. 
Hence, we trivially deduce
$$\sfrac d{dt}\dm\n v^m\dm_2^2+\sfrac12\dm P\Delta v^m\dm_2^2+\dm v^m_t\dm_2^2\leq c\dm \n v^m\dm_2^6\leq c(1+\dm\n v^m\dm_2^2)^2\dm \n v^m\dm_2^2\,.$$
This last is integrated in the following way:
\be\label{DG-IV}\sfrac1{1+\dm\n v^m(0)\dm_2^2}+\intll0t\Big[\sfrac12\sfrac{\dm P\Delta v^m\dm_2^2+\dm v^m_t\dm_2^2}{(1+\dm \n v^m\dm_2^2)^2\hskip-0.15cm}\hskip0.15cm\Big]d\tau\leq \sfrac1{1+\dm\n v^m(t)\dm_2^2} +c\intll0t\dm\n v^m\dm_2^2d\tau\,.\ee Applying H\"older's reverse inequality with ``complementary'' exponents $q=-\frac12$ and $q'=\frac13$, uniformly in $m\in\N$, we get
$$\ba{ll}\displ\Big[\intll0t(\dm P\Delta v^m\dm_2^\frac23+\dm v_t^m\dm_2^\frac23)d\tau\Big]^3\hskip-0.2cm&\displ\leq \Big[\intll0t(1+\dm\n v^m\dm_2^2)d\tau\Big]^2\Big[1+c\intll0t\dm\n v^m\dm_2^2d\tau\Big]\VSE\leq \Big[t+\dm v_0\dm_2^2\Big]^2\Big[1+c\dm v_0\dm^2_2\Big]=: \frak A(t,\dm v_0\dm_2)\,,\mbox{ for all }t>0\,.\ea$$ For any pair $(m,p)$ and for all $t>0$, by means of an integration by parts and Holder's inequality, we obtain $$\ba{ll}\displ\intll0t\hskip-0.08cm\dm \n v^m\hskip-0.07cm-\n v^p\dm_2d\tau\!\leq \!\!\intll0t \hskip-0.08cm\dm P\Delta(v^m\hskip-0.08cm-v^p)\dm_2 ^\frac12
\dm v^m\hskip-0.08cm-v^p\dm_2^\frac12d\tau\hskip-0.3cm&\displ\leq\! \Big[\!\intll0t\hskip-0.08cm \dm P\Delta(v^m\hskip-0.08cm-v^p)\dm_2^\frac23\Big]^\frac34 \hskip-0.05cm\Big[\intll0t\hskip-0.08cm\dm v^m\hskip-0.08cm-v^p\dm_2^2d\tau\Big]^\frac14\VSE\leq2^\frac12 \frak A(t,\dm v_0\dm_2)^\frac1{4}\Big[\intll0t\dm v^m\hskip-0.07cm-v^p\dm_2^2d\tau\Big]^\frac14\,.\ea$$ By virtue of Lemma\,\ref{SCLD},
we obtain the Cauchy condition for $\{ v^m\}$ in $L^1(0,T;J^{1,2}(\OO))$. Via \rf{NSM-II}, the sequence $\{ v^m\}$ is bounded in $L^2(0,T;J^{1,2}(\OO))$. Hence, by interpolation we realize  the Cauchy condition in $L^q(0,T;J^{1,2}(\OO))$, for all $q\in[1,2)$. So that, the sequence admits the strong limit in $L^q(0,T;J^{1,2}(\OO))$, for all $q\in[1,2)$, which coincides with $v\in L^2(0,T;J^{1,2}(\OO))$.  The lemma is proved.
\ep
\begin{coro}\label{CSB}{\it  The sequence $\{v^m\}$ of solutions   to problem \rf{NSM}  converges      to $v$   in $J^{1,2}(\OO)$ uniformly on $t\geq\theta$\,. }\end{coro} \bp By virtue of Lemma\,\ref{SCLD} and Lemma\,\ref{SCG} we have that $\{v^m\}$   strongly converges to $v$   in $J^{1,2}$ a.e. in $t>0$. Without invalidating the thesis, we can assume that the strong convergence holds for $t=\theta$.  We can state the following estimates with bounds ${\frak D}_1$ and ${\frak D}_2$ independent of $m,\,t$ and $v$, and whose values are inessential for our aims;\\\noindent employing    estimates \rf{GNM-I}, \rf{NSM-II} and \rf{NSM-III}   we get
\be\label{CSB-II}\ba{ll}\displ\intll\theta t\dm v^m(\tau)\dm_\infty^2d\tau\hskip-0.2cm&\displ\leq c\Big[\intll\theta t\dm P\Delta v^m(\tau)\dm_2^2d\tau\Big]^\frac12\Big[\intll\theta t\dm \n v^m(\tau)\dm_2^2d\tau\Big]^\frac12=:\frak D_1\,,\\&\hskip0.2cm \mbox{ for all }m\in\N\mbox{ and }t>\theta\,;\ea\ee\noindent employing estimates \rf{NSM-II} and \rf{NSM-III} , we get\be\label{DQ}\ba{ll}\displ\intll\theta t\dm \n v^m(\tau)\dm_2^4d\tau\hskip-0.2cm&\displ\leq \max_{[\theta,t]}\dm \n v^m(\tau)\dm_2^2\intll\theta t\dm \n v^m(\tau)\dm_2^2d\tau=:\frak D_2\,,\\&\mbox{ for all }m\in\N\mbox{ and }t>\theta\,.\ea\ee We denote by  $\frak D:=\max\{\frak D_1,\frak D_2\}$. Set $w:=v^m-v^p$. From system  \rf{NSM} by difference, we deduce
\be\label{CSB-III}\sfrac d{dt}\dm \n w\dm_2^2+\dm P\Delta w\dm_2^2+\dm w_t\dm_2^2=\dm P(v^m\cdot\n w)+P(w\cdot\n v^p)\dm_2^2\,\mbox{ for all }t>\theta\,.\ee Applying H\"older's inequality to the term of the right-hand side, we get
$$\ba{ll}\dm P(v^m\cdot\n w)+P(w\cdot\n v^p)\dm_2^2\hskip-0.2cm&\leq 2(\dm v^m\dm_\infty^2\dm\n w\dm_2^2+\dm w\dm_\infty^2\dm\n v^p\dm_2^2)\VSE\leq c(\dm v^m\dm_\infty^2\dm\n w\dm_2^2+\dm \n w\dm_2\dm P\Delta w\dm_2\dm \n v^p\dm_2^2)\,,\ea$$  where in the last estimate we employed \rf{GNM-I} again. Estimating the right-hand side of \rf{CSB-III} via the last estimate, and employing the Young inequality, we deduce the differential equation
$$\sfrac d{dt}\dm \n w\dm_2^2\leq c\dm \n w\dm_2^2(\dm v^m\dm_\infty^2+\dm \n v^p\dm_2^4)\,,\mbox{ for all }t>\theta\,.$$ Hence, by means of estimates  \rf{CSB-II} and \rf{DQ},  via an  integration, we arrive at 
\be\label{SCG-I}\ba{ll}\dm \n v^m(t)-\n v^p(t)\dm_2\hskip-0.2cm&=\dm \n w(t)\dm_2\leq \hskip0.05cm\exp[\frak D]\hskip0.05cm\dm \n w(\theta)\dm_2\VSE\leq \hskip0.05cm\exp[\frak D]\hskip0.05cm\dm \n v^m(\theta)-\n v^p(\theta)\dm_2\,,\mbox{ for all }m,p\in\N\mbox{ and }t\geq\theta\,.\ea\ee In the case of the $L^2$-norm, for all $t>\theta$, we consider the energy relation related to $w$:
$$\ba{ll}\frac d{dt}\dm w(t)\dm_2^2+2\dm \n w(t)\dm_2^2\hskip-0.2cm&=(w\cdot\n w,v^m)\leq \dm w\dm_6\dm \n w\dm_2\dm v^m\dm_3\leq \dm w\dm_6\dm \n w\dm_2\dm v^m\dm_2^\frac12\dm\n v^m\dm_2^\frac12\VSE\leq c\dm v_0\dm_2^\frac12\dm \n w\dm_2^2\dm \n v^m\dm_2^\frac12\leq c\dm v_0\dm_2^\frac12\dm \n w\dm_2^\frac12\Big[\dm \n v^m\dm_2^2+\dm \n v^p\dm_2^\frac32\dm\n v^m\dm_2^\frac12\Big] ,\ea$$ where we estimated by means of H\"older's inequality, Sobolev inequality and the Gagliardo-Nirenberg inequality.   Taking estimate \rf{NSM-II} into account, an integration on $t> \theta$,   furnishes
\be\label{SCLD-I}\ba{ll}\dm v^m(t)\!-v^p(t)\dm_2^2\hskip-0.3cm&=\dm w(t)\dm_2^2\leq \dm w(\theta)\dm_2^2+c\exp[\frak D]\dm v_0\dm_2^\frac52\hskip0.05cm\dm \n v^m(\theta)-\n v^p(\theta)\dm_2^\frac12\VSE\! \leq \!\dm v^m(\theta)\!-v^p(\theta)\dm_2^2\!+c\exp[\frak D]\dm v_0\dm_2^\frac52 \dm \n v^m(\theta)-\!\n v^p(\theta)\dm_2^\frac12,\VSE\hskip8cm\mbox{\,for\,all\,}m,p\!\in\!\N.\ea\ee  Now, from our assumption of strong convergence of $\{v^m(\theta)\}$ in $J^{1,2}(\OO)$, via estimates \rf{SCG-I} and \rf{SCLD-I},  the thesis of the corollary holds.
\ep
\begin{lemma}\label{GPM}{\sl The sequence $\{v^m\}$  furnished by Theorem\,\ref{NSM-T} enjoys the following estimate:\be\label{DG-III}\intll0t\sfrac1{(1+\dm \n v^m\dm_2^2)^2}\Big|\sfrac d{dt}\dm \n v^m\dm_2^2\Big|d\tau\leq 1+c\dm v_0\dm_2^2\,,
 t>0\,.\ee }\end{lemma}
\bp
From  \rf{A}  we are able to get
\be\label{ALM-III}\ba{ll}\displ\sfrac1{(1+\dm \n v^m\dm_2^2)^2}\Big|\sfrac d{dt}\dm \n v^m\dm_2^2\Big|\hskip-0.2cm&\displ=\Big|\sfrac{\dm P( J_m[v^m]\cdot\n v^m)\dm_2^2}{(1+\dm \n v^m\dm_2^2)^2}-\sfrac{\dm P\Delta v^m\dm_2^2+\dm v^m_t\dm_2^2}{(1+\dm \n v^m\dm_2^2)^2}\Big|\VSE
\leq c\sfrac{\dm\n v^m\dm_2^3\dm P\Delta v^m\dm_2}{(1+\dm \n v^m\dm_2^2)^2}+\sfrac {\dm P\Delta v^m\dm_2^2+\dm v^m_t\dm_2^2}{(1+\dm \n v^m\dm_2^2)^2}\VSE\leq \dm \n v^m\dm_2^2+c\sfrac {\dm P\Delta v^m\dm_2^2+\dm v^m_t\dm_2^2}{(1+\dm \n v^m\dm_2^2)^2}\,,\ea\ee
where, in estimating the right hand-side in \rf{ALM-III}$_1$, we took \rf{ALM-II} into account and applied the Young  inequality. Integrating \rf{ALM-III} on $(0,t)$, via estimate \rf{DG-IV}   and  via  energy relation \rf{NSM-II}, we arrive at \rf{DG-III}. 
\ep
\section{\label{PSEC}Proof  of Theorem\,\ref{CT} }
\subsection{Existence of a weak solution $v$ to problem \rf{NS}}
We start by proving the  existence of a weak solution. 
\begin{lemma}\label{EX-T}{\sl For all $v_0\in J^2(\OO)$ there exists a sequence $\{v^m\}$,   enjoying the following limit properties: \be\label{EST-O}\ba{ll}\mbox{for\,all\,}T>\!0,\hskip-0.3cm &\{v^m\}\!\rightharpoonup \!\hskip0.03cmv\mbox{ in }L^2(0,T;J^{1,2}(\OO))\,,\VSE   \{v^m\}\!\to\! v \mbox{ in }L^2(0,T;L^2(\OO))\cap L^q(0,T;J^{1,2}(\OO)), \,q\!\in\![1,2)\,,\VS&\{v^m\}\!\to\! v\mbox{\,in\,}J^{1,2}(\OO),\mbox{\,for all\,}t\!\in\![0,\theta)-\!J\mbox{\,and\,for\,all\,}t\geq\theta,\mbox{\,with\,}\theta\!\leq\! c\dm v_0\dm_2^4, \ea\ee where the constant $c$ is independent of    $v_0$, the set $J$ has zero Lebesgue measure.  Moreover,  the limit $v$ is a weak solution  to problem \rf{NS} that   enjoys  the  properties:
\be\label{EXT-I}\hskip-0.1cm\ba{c}\displ\dm v(t)\dm_2^2+2\!\intll st\!\dm \n v\dm_2^2d\tau\leq \dm v(s)\dm_2^2,\mbox{\,for\,all\,}t>s,\mbox{\,and\,for\,all\,}s\!\in\! [0,\infty)-I\,,\\\displ \mbox{for\,all\,}s\!\in[0,\infty)-I,\lim_{t\to s^+}\dm v(t)-v(s)\dm_2=0\,,\VS\mbox{for\,all\,}\varphi\!\in\! J^2(\OO),\,  (v(t),\varphi)\mbox{\,is\,a\,continuous\,function\,of }t\,,\ea\ee where the set $I\subseteq J$ has zero Lebesgue measure.}\end{lemma}
 \bp The existence result of $v$   
is achieved following the approach furnished  by Leray in \cite{L}.
Actually, we consider the sequence stated in Theorem\,\ref{NSM-T} and its  limit $v$ stated in Lemma\,\ref{SCLD}, Lemma\,\ref{SCG}. Then the existence of $\theta$ and the related estimate are deduced from Theorem\,\ref{NSM-T}.  For all $t\in [0,T)$, $v$ is  also a weak  limit in $J^{2}(\OO)$ with $(v(t),\varphi)\in C([0,T))$, for all $\varphi\in J^2(\OO)$, that is \rf{EXT-I}$_3$.\par One easily verifies that the limit $v$ is a weak solution.   
\par Property \rf{EST-O}$_3$ is deduced  from the strong convergence stated in Lemma\,\ref{SCG} and the regularity of $v$  for $t>\theta$ from  Corollary\,\ref{CSB}.
\par Estimate \rf{EXT-I}$_1$ is the so called energy inequality in strong form introduced by Leray. In order to deduce \rf{EXT-I}$_1$, we consider an instant $s\geq0$ in which   $\displ\lim_m\dm v^m(s)\dm_2=\dm v(s)\dm_2$. In the case of $s\ne0$, the existence is ensured, almost everywhere in $s\in(0,\theta)$, by Lemma\,\ref{SCLD}, and for all $s\geq\theta$ by Corollary\,\ref{CSB}. We denote by $I$ the set of the possible instants $s$ for which the strong convergence fails to hold. Of course the following inclusion holds: $I\subseteq J$, where $J$ is defined in \rf{EST-O}$_3$. By our arguments, $I\subset (0,\theta)$ and $\mu_1(I)=0$. Then, for all $s\in(0,\infty)-I$, we consider \rf{NSM-II}, and we perform the lower  limit of the left-hand side and the limit of the right-hand side of \rf{NSM-II}. Then one easily achieves the energy inequality for the weak limit $v$ as stated in \rf{EXT-I}. \par Property \rf{EXT-I}$_2$ is an immediate consequence of the continuity expressed in \rf{EXT-I}$_3$ and of \rf{EXT-I}$_1$. \ep 
\begin{coro}\label{SPC}{\sl Assume that for the weak solution given in Lemma\,\ref{EX-T} the energy relation \rf{E-II} holds. Then, for all $T>0$, the sequence $\{\n v^m\}$  converges to $\n v$ in $L^2(0,T;L^2(\OO))$.}\end{coro}
\bp Actually, following argument lines related to the asymptotic behavior of the kinetic energy of a weak solution (see {\it e.g.}\cite{GM,MEC-O,MEC-I,MS,St}) for the sequence $\{v^m\}$ the following identity holds from \rf{NSM-II} for $s=0$:\be\label{ST-I}2\intll0\infty\dm \n v^m(\tau)\dm_2^2d\tau=\dm v_0^m\dm_2^2\,.\ee
Relation \rf{ST-I} and assumption \rf{E-II} allow us to trivially  obtain
$$\intll0\infty\!\!\dm\n v^m(\tau)\dm_2^2\hskip-0.05cm+\dm \n v(\tau)\dm_2^2-2(\n v^m(\tau),\n v(\tau))d\tau =\sfrac12\dm v_0^m\dm_2^2\!+\sfrac12\dm v_0\dm_2^2-2\!\!\intll0\infty\!(\n v^m(\tau),\n v(\tau))d\tau \,.$$ Hence, letting $m\to\infty$ and employing   \rf{E-II} again, we get
$$\lim_m\intll0\infty\dm\n v^m(\tau)-\n v(\tau)\dm_2^2d\tau=\dm v_0\dm_2^2-2\intll0\infty\dm \n v(\tau)\dm_2^2d\tau=0\,.$$The thesis is achieved.
\ep
\subsection{\label{FPC}The  energy equality for the weak solution of  Lemma\,\ref{EX-T}}
 We introduce an auxiliary function $p(\alpha,r)$. To this end, for $\alpha\in(0,1)$, we set
$$\ba{c}  A(\alpha):=\sfrac{\sqrt3}2\tan\alpha\sfrac\pi2+\sfrac12\,,\VS F(\alpha):=\sfrac1{\sqrt3}\sfrac\pi2(1-\alpha)-\sfrac16\log\sfrac{(1+A(\alpha))^2}{1-A(\alpha)+A^2(\alpha)}\,,\VS\mbox{for }r>0\,,\; E(r):=\sfrac1{\sqrt3}\arctan\sfrac{2r -1}{\sqrt 3}+\sfrac16\log\sfrac{(1+r)^2}{1-r+r^2}\,,\mbox{ with }\sfrac d{dr}E(r)=\sfrac1{1+r^3}\,.\ea$$ We have: $A(0)=\frac12$ and $F(0)=\frac12\big(\frac\pi{\sqrt3}-\frac13\log 3\big)>0$. Letting $\alpha\to1^-$,  we get $A(\alpha)\to \infty$. Moreover,  since $\frac1{\sqrt3}\frac\pi2\alpha=\frac1{\sqrt3}\arctan\frac{2A(\alpha)-1}{\sqrt 3}$, we realize that $F(\alpha)=\sfrac12\sfrac1{\sqrt 3}-E(A(\alpha))\,.$ Hence, we get $$F'(\alpha)=-\sfrac1{1+A^3(\alpha)}A'(\alpha)=-\sfrac1{1+A^3(\alpha)}\sfrac {\sqrt3}2\frac\pi2(\cos\alpha\ssfrac\pi2)^{-2}<0\,,\mbox{ for all }\alpha\in(0,1)\,,$$  then  $F(\alpha)$ decreases monotonically to $0$. We define 
\be\label{COF}p(\alpha,r):=\left\{\ba{ll}\hskip1cm1\,, &\mbox{if }r\leq A(\alpha)\,,\VS
\frac{\frac1{\sqrt3}\frac\pi2-E(r)}{F(\alpha)}\,,&\mbox{if }r\geq A(\alpha)\,.\ea\right.\ee  
Hence, $p(\alpha,r)$ is a Lipschitz function with  
\be\label{DP}\sfrac {d\hskip0.05cm p(\alpha,r)}{dr}=\left\{\ba{ll}\hskip1cm 0\,,&\mbox{if }r\leq A(\alpha)\,,\VS-\sfrac1{F(\alpha)}\sfrac 1{1+r^3\hskip-0.2cm}\hskip0.2cm\,,&\mbox{if }r>A(\alpha)\,.\ea\right.\ee
Thus, for all $\alpha\in(0,1)$,   $p(\alpha,r)$ is positive, monotone decreasing  on $(A(\alpha),\infty)$ with limit $0$ at infinity. 
     \par Recalling \rf{EST-O}$_3$, we denote by  $$\mathcal T\!:=\{s\geq 0: \{v^m(s,x)\} \mbox{ is strongly convergent to }v\mbox{ in }J^{1,2}(\OO)\}\,,\mbox{ and }J\!:=(0,\infty)-\mathcal T.$$ \par  Let $s,t\in \mathcal T$. Then there exists $\alpha_0$ such that \be\label{ABV}\dm \n v^m(t)\dm_2^2<A(\alpha) \mbox{ and }\dm\n v^m(s)\dm_2^2<A(\alpha)\,,\mbox{ for all }\alpha\in (\alpha_0,1)\mbox{ and }m\in\N\,.\ee
   We set $\rho_m(t):=\dm \n v^m(t)\dm_2^2$ \,, $e_m(t):=\dm v^m(t)\dm_2^2$\, and, for $\alpha\in(\alpha_0,1)$,
 $$J(\alpha,m):=\{\tau\in(s,t):\rho_m(\tau)\in(A(\alpha),A^m]\}\,,$$ where $A^m:=\displ\max_{[s,t]} \rho_m$\,.\vskip0.1cm\par {\bf i.} If $A^m\leq A(\alpha)$ holds, then $J(\alpha,m)$ is an empty set.\par {\bf ii.} If $A^m>A(\alpha)$ holds,  since $\max\{\rho_m(s),\rho_m(t)\}<A(\alpha)$\,, then one gets $J(\alpha,m)\ne\emptyset$\,.\par In the case {\bf ii.},  the intersection between the straight line $(\sigma,A(\alpha))$  and the regular curve $(\sigma,\rho_m(\sigma))$, $\sigma\in(s,t)$, by the continuity of $\rho_m(\sigma)$, detects the endpoints of open not empty intervals. In this connection, we say that $\ov\tau\in J(\alpha,m)$ is a left endpoint if there exists a neighborhood $(-\ov\delta+\ov\tau,\ov\tau+\ov\delta_1)$ such that $\rho_m(\tau)\leq A(\alpha)$ for $\tau\in(-\ov\delta+\ov\tau,\ov\tau]$ and $\rho_m(\tau)>A(\alpha)$ for $\tau\in (\ov\tau,\ov\tau+\ov\delta_1)$. Analogously, we say that $\ov{\ov\tau}\in J(m,\alpha)$ is a right endpoints if there exists a neighborhood $(-\ov{\ov\delta}_2+\ov{\ov\tau},\ov{\ov\tau}+\ov{\ov\delta})$ such that $\rho_m(\tau)>A(\alpha)$ for $\tau\in (-\ov{\ov\delta}_2+\ov{\ov\tau},\ov{\ov\tau})$ and $\rho_m(\tau)\leq A(\alpha)$ for $\tau\in [\ov{\ov\tau},\ov{\ov\tau}+\ov{\ov\delta})$. Actually, by virtue of \rf{ABV} related to $s$ and by continuity of $\rho_m(\tau)$ there exists $\ov\tau>s$ such that $\rho_m(\ov\tau)=A(\alpha)$ and $\rho_m(\tau)>A(\alpha)$ for some open right-neighborhood $U^+(\ov \tau)$. Analogously, by virtue of \rf{ABV} related to $t$ and by the continuity of $\rho_m(\tau)$, there exists $\ov{\ov \tau}<t$ such that $\rho_m(\ov{\ov \tau})=A(\alpha)$ and $\rho_m(\tau)>A(\alpha)$ in some open left-neighborhood $U^-({\ov\tau})$.\par  Hence, if the intersection between the straight line  $(\sigma,A(\alpha))$ and the curve $(\sigma,\rho_m(\sigma))$ holds in different points, iterating the previous argument, we state the existence of distinct and successive points of the kind ${\ov\tau}$ and $\ov{\ov\tau}$ as endpoints of a not empty interval $U\subseteq J(\alpha,m)$. Being  $U\ne\emptyset$ and open, $J(\alpha,m)$ is at most the union of a countable family of pairwise disjoint intervals $(s_h,t_h)$ with $\rho(s_h)=A(\alpha)$ and $\rho(t_h)=A(\alpha)$. We denote by $\N(\alpha,m)$ the set of indexes $h$, for which we know that the cardinality is at most $\aleph_0$.\par
 We multiply the equation \rf{NSM}$_1$ by $v^mp(\alpha,\rho_m)$ and we integrate by parts on $(s,t)\times \OO$. Then we get
\be\label{EM-I}e_m(t)+2\intll stp(\alpha,\rho_m)\rho_m(\tau)d\tau+\widetilde I(\alpha,m)=e_m(s)\,,\mbox{ for all }m\in\N\,,\ee where via \rf{ABV}  we took $p(\alpha,\dm \n v^m(t)\dm_2^2)=p(\alpha,\dm \n v^m(s)\dm_2^2)=1$ into account. Recalling \rf{DP}, 
we set \be\label{DI}\widetilde I(\alpha,m):=\sfrac1{F(\alpha)}\!\intl{J(\alpha,m)}  \sfrac{e_m(\tau)\dot\rho_m(\tau)}{1+\rho^3_m(\tau)}d\tau=:\sfrac{I(\alpha,m)}{F(\alpha)} \,.\ee       In order to give an estimate of the measure of $J(\alpha ,m)\subset(s,t)$, we consider the following $$\mbox{$\underset {h\in \N(\alpha,m)}\sum$}\!\!(t_h-s_h)A(\alpha)=|J(\alpha ,m)|A(\alpha)\leq \!\!\intl{J(\alpha ,m)}\!\!\!\rho_m(\tau)d\tau\leq \intll st\rho_m(\tau)d\tau\leq\sfrac12\dm v_0\dm_2^2\,,$$  where we took the energy relation \rf{NSM-III} into account. Hence, we get
\be\label{MJ}|J(\alpha,m)|\leq c\dm v_0\dm_2^2[ A(\alpha)]^{-1}\,,\mbox{ for all }m\in\N\,.\ee
 Letting $m\to\infty$ in \rf{EM-I}, via the limit property \rf{WC-I}, we arrive at
\be\label{PC}e(t)+2\intll st\rho(\tau)p(\alpha,\rho)d\tau+\lim_m\widetilde I(\alpha,m)=e(s)\,,\mbox{ for all }\alpha\in(\alpha_0,1)\,,\ee and, employing \rf{WC-IO}, \be\label{ISE}e(t)+2\intll st\rho(\tau)d\tau+\lim_{\alpha\to1^-}\lim_m\widetilde I(\alpha,m)=e(s)\,.\ee 
Now, we estimate $I(\alpha,m)$ defined in \rf{DI}. Since for all $\alpha\in(\alpha_0,1)$ and $m\in\N$ there exists $\N(\alpha,m)\subseteq\N$ such that  $J(\alpha,m)=\underset{h\in \N(\alpha,m)}\cup (s_h,t_h)$ with $(s_h,t_h)\cap(s_{h'},t_{h'})=\emptyset$ for all $h\ne h'$,  we get
\be\label{SC}I(\alpha,m)=\mbox{$\underset{h\in\N(\alpha,m)}\sum$}\intll{s_h}{t_h}\sfrac{e_m(\tau)\dot \rho_m(\tau)}{1+\rho_m^3(\tau)}d\tau\,.\ee 
For all index $h\in \N(\alpha,m)$, an integration by parts on $(s_h,t_h)$ furnishes
\be\label{IBP} \ba{ll}\displ\intll{s_h}{t_h}\!\!\ssfrac{e_m(\tau)\dot \rho_m(\tau)}{1+\rho_m^3(\tau)}d\tau\hskip-0.3cm&\displ=\ssfrac{A(\alpha)}{1+A^3(\alpha)\hskip-0.1cm}\big[e_m(t_h)-e_m(s_h)\big]\!-\!\!\intll {s_h}{t_h}\!\!\ssfrac{\dot e_m(\tau)\rho_m(\tau)\hskip-0.1cm}{1+\rho_m^3(\tau)}\hskip0.05cmd\tau\!+3\!\!\intll {s_h}{t_h}\!\!\ssfrac{e_m(\tau)\rho_m^3(\tau)\dot\rho_m(\tau)\hskip-0.1cm}{(1+\rho_m^3(\tau))^2}d\tau\\&\displ= -2\ssfrac{A(\alpha)}{1+A^3(\alpha)}\intll{s_h}{t_h}\!\!\rho_m(\tau)d\tau+2\!\intll {s_h}{t_h}\!\ssfrac{\rho_m^2(\tau)\hskip-0.1cm}{1+\rho_m^3(\tau)}\hskip0.1cmd\tau\\&\displ\hskip6cm+3\!\intll {s_h}{t_h}\!e_m(\tau)\dot\rho_m(\tau)\ssfrac{\rho_m^3(\tau)}{(1+\rho_m^3(\tau))^2\hskip-0.1cm}\hskip0.1cmd\tau,\ea\ee where we took the formula \rf{NSM-II} into account.
The following identity holds: $\frac1{1+y^3}-\frac1{(1+y^3)^2} =\frac{y^3}{(1+y^3)^2}$. In the last integral in \rf{IBP}, we  substitute  the quoted identity with $y:=\rho_m(\tau)$. Hence,  for any $h$-term,   we also have
\be\label{TRS}2\intll{s_h}{t_h}\sfrac{e_m\dot\rho_m(\tau)}{1+\rho_m^3(\tau)}d\tau=2\sfrac{A(\alpha)}{1+A^3(\alpha)}\intll{s_h}{t_h}\rho_md\tau-2\intll{s_h}{t_h}\sfrac{\rho_m^2(\tau)}{1+\rho_m^3(\tau)}d\tau+3\intll{s_h}{t_h}\sfrac{e_m\dot\rho_m}{(1+\rho_m^3)^2}d\tau\,.\ee
The  first term on the right-hand side of \rf{TRS}, for $h\in \N(\alpha,m)$, form a  series  with non negative terms and it is well posed being
\be\label{BD-I}\mbox{$\underset{h\in\N(\alpha,m)}\sum$}2\intll{s_h}{t_h}\rho_m(\tau)d\tau\leq 2\intll st\rho_m(\tau)d\tau\leq \dm v_0\dm_2^2\,.\ee
The second term on the right hand side of \rf{TRS}, for $h\in \N(\alpha,m)$, form a  series that converges being the series with non negative  terms finite
\be\label{BD-II}\mbox{$\underset{h\in\N(\alpha,m)}\sum$}\!\intll {s_h}{t_h}\!\sfrac{\rho_m^2(\tau)\hskip-0.1cm}{1+\rho_m^3(\tau)}\hskip0.1cmd\tau\leq [1+A(\alpha)]^{-1}|J(\alpha,m)|\leq c\dm v_0\dm_2^2[1+A(\alpha)]^{-2}\,,\ee where  we employed \rf{MJ}. Recalling   the energy relation \rf{NSM-II} for the term $e_m(t)$ and estimate \rf{DG-III}, the last term on the right-hand side of \rf{TRS},  for $h\in \N(\alpha,m)$, forms a  series that   converges being bounded by an   absolute convergent series:   
\be\label{BD-III}\ba{ll}\displ\Big|\mbox{$\underset{h\in\N(\alpha,m)}\sum$}\!\intll {s_h}{t_h}\!\!\sfrac{e_m(\tau)\dot\rho_m(\tau)\hskip-0.1cm}{(1+\rho_m^3(\tau))^2}\hskip0.1cmd\tau\Big|\hskip-0.2cm&\displ\leq \mbox{$\underset{h\in\N(\alpha,m)}\sum$}\!\intll {s_h}{t_h}\!\!\sfrac{e_m(\tau)|\dot\rho_m(\tau)|\hskip-0.1cm}{(1+\rho_m^3(\tau))^2}\hskip0.1cmd\tau\\&\displ\leq c[1+A(\alpha,m)]^{-2}\dm v_0\dm_2^2\!\intll {s}{t}\!\!\sfrac{|\dot\rho_m(\tau)|\hskip-0.1cm}{(1+\rho_m^2(\tau))}\hskip0.1cmd\tau\VSE\displ\leq c[1+A(\alpha,m)]^{-2}(\dm v_0\dm_2^2+\dm v_0\dm_2^4) \,.\ea\ee 
Hence, we write
$$\ba{ll}I(\alpha,m)\hskip-0.2cm&\displ=2\sfrac{A(\alpha)}{1+A^3(\alpha)}\mbox{\Large$\underset{h\in\N(\alpha,m)}\sum$}\intll{s_h}{t_h}\!\rho_md\tau-2\!\!\mbox{\Large$\underset{h\in\N(\alpha,m)}\sum$}\intll{s_h}{t_h}\!\sfrac{\rho_m^2(\tau)}{1+\rho_m^3(\tau)}d\tau+3\!\!\mbox{\Large$\underset{h\in\N(\alpha,m)}\sum$}\intll{s_h}{t_h}\!\sfrac{e_m\dot\rho_m}{(1+\rho_m^3)^2}d\tau\VSE=:I_1(\alpha,m)+I_2(\alpha,m)+I_3(\alpha,m)\,.\ea$$
Now, we prove that 
\be\label{IISE}\lim_{\alpha\to1^-}\lim_m\widetilde I(\alpha,m)\leq0\,.\ee
Actually, for all $\alpha\in(\alpha_0,1)$ the sequences $\{I_i(\alpha,m)\}$, $i=1,2,3$, admit a bound, that is  \rf{BD-I}, \rf{BD-II} and \rf{BD-III}, respectively. 
Hence, being for all $\alpha\in(\alpha_0,1)$ the sequences $\{I_i(\alpha,m)\}$, $i=1,2,3$ bounded, they admit a subsequence that converges to $L_i(\alpha)$, $i=1,2,3$,  and the limits admit the same bound, that is, for $c$ independent of $\alpha$, we get
$$L_1(\alpha)+L_2(\alpha)+L_3(\alpha)\leq c\sfrac{\dm v_0\dm_2^2+\dm v_0\dm_2^4}{1+A^2(\alpha)}\,,$$ where for $I_1(\alpha,m)$ we took the coefficient $\frac{A(\alpha)}{1+A^3(\alpha)}$ into account too. Since by virtue of \rf{PC}, for all $\alpha\in(\alpha_0,1)$, the sequence  $I(\alpha,m)$ is convergent, and the convergence also holds to the same limit  for any  subsequence, we get 
$$\lim_mI(\alpha,m)\leq c  \sfrac{\dm v_0\dm_2^2+\dm v_0\dm_2^4}{1+A^2(\alpha)}\,,\mbox{ for all }\alpha\in(\alpha_0,1)\,.$$ Multiplying by $(F(\alpha))^{-1}$ and letting $\alpha\to1^-$, via a suitable constant $c$, we get
$$\lim_{\alpha\to1^-}\!\lim_m\widetilde I(\alpha,m)=\!\lim_{\alpha\to1^-}F^{-1}(\alpha)\lim_mI(\alpha,m)\leq\! \lim_{\alpha\to1^-}\!\sfrac{c}{\big[\frac1{\sqrt3}\frac\pi2(1-\alpha)-\frac16\log\frac{(1+A(\alpha))^2}{1-A(\alpha)+A^2(\alpha)}\big]
A^2(\alpha)}=0\,,$$
being $A(\alpha)=\frac{\sqrt3}2\tan\alpha\frac\pi2+\frac12$ and
$$\lim_{\alpha\to1^-}A(\alpha)(1-\alpha)=\sfrac{\sqrt3}\pi\mbox{\; and\; }\lim_{\alpha\to1^-}A(\alpha)\log\sfrac{(1+A(\alpha))^2}{1-A(\alpha)+A^2(\alpha)} =3\,.$$ We have proved estimate \rf{IISE}.
As a consequence of \rf{ISE} and \rf{IISE}, we realize that
$$\dm v(s)\dm_2^2-2\intll st\dm \n v(\tau)\dm_2^2d\tau-\dm v(t)\dm_2^2\leq 0\,.$$  Hence, via \rf{EXT-I}$_1$ we conclude the proof related to the energy equality, provided that $t,s\in\mathcal T$, that is \be\label{PEE}\dm v(t)\dm_2^2+2\intll st\dm\n v(\tau)\dm_2^2d\tau=\dm v(s)\dm_2^2\,.\ee The case of $s=0$ is an easy consequence of the fact that there exists $\{s_h\}\subset\mathcal T$ and converging to 0 such that $\displ \lim_{s_h}\dm v(s_h)-v_0\dm_2=0$. Hence, evaluating \rf{PEE} on $(s_h,t)$ and then letting $s_h\to0$, we conclude the proof of \rf{E-I}.  If $I-J\ne\emptyset$,  the same arguments hold considering $t\in I-J$. Actually, for such a $t$, on the right side we have the continuity in $L^2$-norm of the weak solution, that is \rf{EXT-I}$_2$. Hence, considering $\{t_n\}\subset J$ that converges to $t$ from the right, we get
$$\dm v(t)\dm_2^2+2\intll st\dm\n v(\tau)\dm_2^2d\tau=\lim_n\Big[\dm v(t_n)\dm_2^2+2\intll s{t_n}\dm\n v(\tau)\dm_2^2d\tau\Big]=\dm v(0)\dm_2^2\,.$$ \par In order to prove formula \rf{E-II}, it is enough to consider the known results on the asymptotic behavior of the kinetic energy (in this connection see {\it e.g} \cite{GM,MEC-O,MEC-I,MS,St}). These results ensure that in the limit for $t\to\infty$ we arrive at \rf{E-II}.\section{Appendix}
In this section, we give a possible proof  of Theorem\,\ref{NSM-T}, that, as  said, is a result already known in literature.
The proof is based on the results concerning the IBVP for Stokes problem:
\be\label{ST}\ba{c}u_t+\n\pi_u=\Delta u+f\,,\;\n \cdot u=0\,\mbox{ in }(0,T)\times\OO\,,\VS v=0\mbox{ on }(0,T)\times\po\,, \;u=u_0\mbox{ on }\{0\}\times\OO\,.\ea\ee
We recall the following result due to Solonnikov.
\begin{tho}\label{TES}{\sl Let $u_0\in J^{1,2}(\OO)$ and $f\in L^2(0,T;L^2(\OO))$. Then there exists a unique solution to  problem \rf{ST} such that
\be\label{SEL}\ba{l}\displ\dm u(t)\dm_{1,2}^2+\intll0t \Big[\dm \n u(\tau)\dm_2^2+\dm P\Delta u(\tau)\dm_2^2+\dm u_\tau(\tau)\dm_2^2\Big]d\tau\\\displ\hskip6cm\leq c\dm u_0\dm_{1,2}^2+c(T)\intll0T\dm f(\tau)\dm_2^2d\tau\,,\mbox{ for all }T>0\,,\ea\ee where $c(T)\in[0,\infty)$ is independent of $u$\,. In particular, $u\in C([0,T);J^{1,2}(\OO))$ holds.}\end{tho}
\bp We refer the reader to the paper  by  Solonnikov \cite{Sl}, and to more recent contributes furnished  in \cite{MS-I,MS-II}.\ep
\begin{lemma}\label{SAS}{\sl 
Let $a >0$ and $c>0$. Let $\{\xi_p\}$ be a nonnegative sequence of real numbers such that:
\be\label{SAS-I} 
\xi_p\leq a+c\,\xi_{p-1}^2, \mbox{ for }p\geq2\,.
\ee
Assume $1-4ca>0$, $\xi_0=0$, and  $\xi_1\leq \xi$, where $\xi$ is the minimum solution of algebraic equation $c\xi^2-\xi+a=0$. Then $\xi_{p-1}\leq \xi$ for all $p\in \N$.}
\end{lemma}
\bp See Lemma\,10.2 in \cite{Sl}\,.\ep
In order to state the result of existence and regularity for solutions to problem \rf{NSM}, we  define the following one:
\be\label{NSM-A}\ba{cc}{v^m_p}_{\!\!t}+ \n\pi^m_{p}-\Delta v^m_{p}=-\mathbb J_m[v^{m}_{p-1}]\cdot\n v^m_{p-1}\,,&\n\cdot v^m_p=0\,,\mbox{ on }(0,T)\times \OO\,,\VS v^m_p=0\mbox{ on }(0,T)\times\po\,,& v^m_p=v^m_0\mbox{ on }\{0\}\times\OO\,.\ea\ee
Hence, we  apply the method of successive approximations in order to obtain a solution $\{(v^m,\pi_{v^m})\}$ defined on some $(0,T(m))$. For the sake of simplicity, we omit the apex $m$ related to the solution of \rf{NSM-A} and we preserve the index $m$ for the mollifier and we spotlight the dependence on $m$ of  the constants that appear in the estimates. 
\par We set for all $t>0$ $$E_p(t):=\dm v_p(t)\dm_{1,2}^2+\intll0t\Big[\dm {v_p}_\tau(\tau)\dm_2^2+\dm \n v_p(\tau)\dm_2^2+\dm P\Delta v_p(\tau)\dm_2^2\Big]d\tau\,.$$
\begin{lemma}\label{LA-I}{For all $m\in\N$, there exist a $T>0$, such that  $1-4T^\frac14c(m)\dm v_0^m\dm_{1,2}^2>0$, and a sequence $\{(v_p,\pi_{v_p})\}$ such that \be\label{A-I} E_p(T)\leq \sfrac{2\dm v_0^m\dm_{1,2}^2}{1+(1-4T^\frac14c(m)\dm v_0^m\dm_{1,2}^2)^\frac12\hskip-0.2cm}\hskip0.1cm=:\xi\,,\mbox{ for all }p\in\N\,.\ee}\end{lemma}
\bp We consider the sequence of solutions $\{(v_p,\pi_{v_p})\}$ where, for $p=1$,  $(v_1,\pi_{v_1})$ is the solution to problem \rf{ST} with $f=0$ and initial datum $v_0^m$, and, for $p>1$, $(v_p,\pi_{v_p})$ is the solution to problem \rf{ST} with initial datum $v^m_0$ and force data $f=\mathbb J_m[v_{p-1}]\cdot \n v_{p-1}$. For all $p\in\N$, the existence is ensured by Theorem\,\ref{TES} provided that $v_{p-1}\in L^2(0,T;L^2(\OO))$. Actually, as consequence of the energy relation for $v_1$, that is 
$$\dm v_1(t)\dm_2^2+2\intll0T\dm\n v_1(\tau)\dm_2^2d\tau\leq \dm v_0^m\dm_2^2\,,$$
 and of the following identity
$$\intll0t\dm {v_1}_\tau(\tau)-P\Delta v_1(\tau)\dm^2_2d\tau=\intll0t\Big[\dm {v_1}_\tau\dm_2^2+\dm P\Delta v_1(\tau)\dm_2^2+\sfrac d{d\tau}\dm \n v_1(\tau)\dm_2^2\Big]d\tau=0\,,$$ we get
$$E_1(t)=\dm v_0^m\dm_{1,2}^2\,,\mbox{ for all }t>0\,.$$
In the case of $p>1$, assuming   $E_{p-1}(t)<\infty$,  we prove that 
\be\label{IT}E_p(t)\leq \dm v^m_0\dm_{1,2}^2+c(m)t^\frac14 (E_{p-1}(t))^2\ee
To this end, we firstly obtain the energy relation. Multiplying  equation \rf{NSM-A} by $v_p$ and integrating by parts, we get
$$\dm v_p(t)\dm_2^2+2\intll0t\dm \n v_p(\tau)\dm_2^2d\tau=\dm v_0^m\dm_2^2+\intll0t(\mathbb J_m[v_{p-1}]\cdot \n v_p,v_{p-1})d\tau\,.$$
Employing   H\"older's inequality in the right-hand side, we easily arrive at
$$\ba{ll}\displ\dm v_p(t)\dm_2^2+\intll0t\dm \n v_p(\tau)\dm_2^2d\tau-\dm v_0^m\dm_2^2\hskip-0.2cm&\displ\leq\intll0t\dm\mathbb J_m[v_{p-1}]\dm_4^2\dm v_{p-1}\dm_4^2d\tau\leq c(m)E_{p-1}(t)\intll0t\dm v_{p-1}(\tau)\dm_4^2d\tau\VSE\hskip-2cm\leq c(m)E_{p-1}(t)\intll0t\dm v_{p-1}(\tau)\dm_2^\frac12\dm \n v_{p-1}(\tau)\dm_2^\frac32d\tau\leq c(m)t^\frac14(E_{p-1}(t))^2\,.\ea$$
Estimating the right-hand side, we employed the Gagliardo-Nirenmberg inequality for $\dm v_{p-1}\dm_4$ and  H\"older's inequality for the time integral. Thus, we arrive at
\be\label{EI}\dm v_p(t)\dm_2^2+2\intll0t\dm \n v_p(\tau)\dm_2^2d\tau\leq\dm v_0^m\dm_2^2+c(m)t^\frac14 (E_{p-1}(t))^2\,.\ee Now, we consider the following identity:
$$\ba{ll}\displ\intll0t\dm {v_p}_\tau(\tau)-P\Delta v_p(\tau)\dm^2_2d\tau\hskip-0.2cm&\displ=\intll0t\Big[\dm {v_p}_\tau\dm_2^2+\dm P\Delta v_p(\tau)\dm_2^2+\sfrac d{d\tau}\dm \n v_p(\tau)\dm_2^2\Big]d\tau\VSE=\intll0t\dm\mathbb J_m[v_{p-1}]\cdot\n v_{p-1}\dm_2^2\leq \intll0t\dm \mathbb J_m[v_{p-1}(\tau)\dm_\infty^2\dm \n v_{p-1}(\tau)\dm_2^2d\tau\VSE \leq c(m)E_{p-1}^\frac54(t)\intll0t \dm \n v_{p-1}(\tau)\dm_2^\frac32d\tau\leq c(m)t^\frac14( E_{p-1}(t))^2\,.\ea$$ Hence, we deduce
$$\dm \n v_p(t)\dm_2^2+\intll0t\Big[\dm v_\tau(\tau)\dm_2^2+\dm P\Delta v_p(\tau)\dm_2^2\Big]d\tau\leq \dm \n v^m_0\dm_2^2+ c(m)t^\frac14(E_{p-1}(t))^2\,.$$
 In particular, the last estimate   and \rf{EI} furnish
$$E_p(t)\leq \dm \n v_0^m\dm_{1,2}^2+c(m)t^\frac14 (E_{p-1}(t))^2\,.$$ Hence,  we arrived at \rf{IT}.
Assuming $T>0$ such that $1-4c(m)\dm v_0^m\dm_{1,2}^2T^\frac14>0$,  by virtue of Lemma\,\ref{SAS}, we get  $$E_p(t)\leq \xi,\,\mbox{ for all }t\in[0,T)\,,$$ where $\xi$ is the minimum root of the algebraic equation
$$T^\frac14c(m)\xi^2-\xi+\dm v_0^m\dm_{1,2}^2\,.$$
The lemma is proved.
\ep
\begin{lemma}\label{LA-II}{\sl The  sequence $\{(v_p,\pi_{v_p})\}$ stated in Lemma\,\ref{LA-I} converges to a limit $(v^m,\pi_{v^m})$ solution to problem \rf{NSM}, and enjoying the properties
\be\label{A-IV}\ba{l} \dm v^m(t)\dm_{1,2}^2\leq \xi\,,\mbox{ for all }t\in(0,T)\,,\VS v^m\in C([0,T;J^{1,2}(\OO))\cap L^2(0,T;W^{2,2}(\OO))\,, \VS v^m_t,\n\pi_{v^m}\in L^2(0,T;L^2(\OO))\,.\ea\ee}\end{lemma} \bp We set the right-hand side of \rf{A-I}
equal to $\xi$.  We initially prove that the sequence  $\{(v_p,\pi_{v_p})\}$  converges  strongly  to a limit $(v^m,\pi_{v^m})$ in $ L^2(0,T;J^{1,2}(\OO))$.
For all $p>1$, we consider the difference $w_p:=v_{p}-v_{p-1}$ and $\pi_{w_p}:=\pi_{v_{p}}-\pi_{v_{p-1}}$.   
After writing the system for $\{(w_p,\pi_{w_p})\}$, we derive the energy relation, initially for index $p=2$:
$$\dm w_2(t)\dm_2^2+2\intll0t\dm\n w_2(\tau)\dm_2^2d\tau=\intll0t(\mathbb J_m[v_{1}]\cdot\n w_{2},v_1)d\tau\,.$$  Applying   H\"older's  and then the Young inequality on the right-hand side, recalling estimate \rf{A-I} and $\dm \mathbb J_m(a)\dm_q\leq \dm a\dm_q$, we easily obtain
$$\dm w_2(t)\dm_2^2+\intll0t\dm\n w_2(\tau)\dm_2^2d\tau\leq \intll0t\dm \mathbb J_m[v_1]\dm_4^2\dm v_1\dm_4^2d\tau\leq \xi^\frac54\intll0t\dm \n v_1\dm_2^\frac32d\tau\,.$$ In the estimate we also employed the Gagliardo-Nirenberg inequality for $\dm v_1\dm_4$\,.  Thus, we arrive at
$$E(w_{2},t):=\dm w_{2}(t)\dm_2^2+\intll0t\dm\n w_{2}(\tau)\dm_2^2\leq c\hskip0.05cm\xi^2t^\frac14\,.$$   Now,   assuming  that 
$$E(w_{p},T):=\dm w_{p}(T)\dm_2^2+\intll0T\dm\n w_{p}(\tau)\dm_2^2\leq \big[c\hskip0.05cm\xi^2\big]^{p-1}T^\frac{p-1}4\,,$$ by the same arguments of the case $p=2$,   we get
$$\dm w_{p+1}(t)\dm_2^2+2\intll0t\dm\n w_{p+1}(\tau)\dm_2^2d\tau=\intll0t\Big[(\mathbb J_m[w_{p}]\cdot\n w_{p+1},v_p)+(\mathbb J_m[v_{p-1}]\cdot\n w_{p+1}, w_{p} )\Big]d\tau\,.$$ Applying   H\"older's  and then the Young inequality on the right-hand side, we obtain
$$\dm w_{p+1}(t)\dm_2^2+\intll0t\dm\n w_{p+1}(\tau)\dm_2^2\leq c \intll0t\dm w_{p}\dm_4^2(\dm v_p\dm_4^2+\dm v_{p-1}\dm_4^2)d\tau\leq (c\hskip0.05cm\xi)^{2p}t^\frac p4\,.$$
We fix $T^\frac14$ in such a way the series of general term $\xi^{2p}t^\frac p4$ is convergent. Hence, by making use of an usual argument and  setting 
$$v_\ell:=\mbox{$\overset \ell{\underset {p\geq2}\sum}$}(v_{p}-v_{p-1})\,,$$ we realize that the sequence $\{v_\ell\}$  converges to $v^m$ in $L^2(0,T;J^{1,2}(\OO))$.
On the other hand via the bound \rf{A-I} the whole sequence $\{v_p\}$ is weakly convergent in the metric detected by $E_p(T)$. Actually, if there exists a subsequence, the subsequence and its complement in $\{v_p\}$ should be convergent to two different limits. But, this is not possible, as the two different limits coincide   with the limit of $\{v^m\}$ in $L^2(0,T;J^{1,2}(\OO))$. One easily verifies that $(v^m,\pi_{v^m})$ is a solution to problem \rf{NSM} on $(0,T)$.
\ep
\begin{lemma}\label{A-P}{\sl The solution $(v^m,\pi_{v^m})$ to problem \rf{NSM} stated in Lemma\,\ref{LA-II} can be extended to all $T>0$ in such a way that \rf{A-IV} holds for all $m\in\N$\,.}\end{lemma}
\bp We state two bounds for the solution $(v^m,\pi_{v^m})$. The first one is related to the energy relation. Multiplying    equation \rf{NSM} by $v^m$, and integrating by parts, we get
\be\label{ERM}\dm v^m(t)\dm_2^2+2\intll0t\dm \n v^m(\tau)\dm_2^2d\tau=\dm v_0^m\dm_2^2\,,\mbox{ for all }t\in[0,T)\,.\ee The energy relation \rf{ERM} is independent of  $T$.
Analogously, by considering the projection of  equation \rf{NSM}$_1$ and the related norm in $L^2(\OO)$, an integration gets 
$$\dm \n v^m(t)\dm_2^2+2\intll0t\big[\dm P\Delta v^m(\tau)\dm_2^2+\dm v^m_\tau(\tau)\dm_2^2\big]d\tau= \dm \n v^m_0\dm_2^2+2\intll0t \dm\mathbb J_m[v^m(\tau)]\cdot\n v^m(\tau)\dm_2^2d\tau\,.$$ Applying   H\"older's inequality on the right-hand side and the properties of the mollifier, we deduce the estimate
\be\label{GRM}\dm \n v^m(t)\dm_2^2\!+2\!\!\intll0t\!\!\big[\dm P\Delta v^m(\tau)\dm_2^2\!+\!\dm v^m_\tau(\tau)\dm_2^2\big]d\tau\leq\! \dm \n v^m_0\dm_2^2\!+2c(m)\dm v_0^m\dm_2^4,\mbox{\,for\,all\,}t\!\in\![0,T),\ee
where we took the energy relation \rf{ERM} for $v^m$ into account.
The previous bounds ensure that $v^m(T,x)\in J^{1,2}(\OO)$. Hence, employing  Lemma\,\ref{LA-II}, we  state an extension  on $(0,T_1)\supset (0,T)$. The extension  enjoys  the same bounds \rf{ERM}-\rf{GRM}, that is $v^m(T_1,x)\in J^{1,2}(\OO)$  with the bounds  independent of $T_1$.
In a finite number of steps, depending only on $m\in\N$, we get the existence of  instant $\theta_m \leq c\dm v_0\dm_2^4$ stated in \rf{ITH}, that is $\theta_m$ such that $\dm \n v(\theta_m)\dm_2\leq (2c\dm v_0\dm_2^2)^{-1}$. Hence, by virtue of Lemma\,\ref{LA-II} we achieve the existence of a solution $(v^m,\pi_{v^m})$ on some interval $(\theta_m,\theta_m+T(\theta_m))$. On such an interval, we apply the estimate \rf{GEX}: $$\sfrac d{dt}\dm \n v^m\dm_2^2+\dm P\Delta v^m\dm_2^2+\dm v^m_t\dm_2^2<\dm\n v^m\dm_2^4 \dm v_0\dm_2^{-2}(2c\dm v_0\dm_2^2\dm \n v^m\dm_2^2-1)_{\big|t=\theta_m}\leq 0\,,$$ where in this last $c$ is a constant independent of $m\in\N$. Hence, we  state that $T(\theta_m)=\infty$ for all $m\in\N$. The existence claimed in
Theorem\,\ref{NSM-T} is proved.
\ep
\vskip0.2cm
{\bf Declaration:\vskip0.1cm\par Conflict of interests} The authors declare no conflict of interest.

 \end{document}